\documentclass[manuscript]{aastex}
\usepackage{epsfig}
\usepackage{latexsym}
\usepackage{amsmath}
\newtheorem{theorem}{Theorem}[section]

\newcommand{\x}{\mathcal S}

\newcommand{\T}{\mathcal T}

\newcommand{\PP}{{\mathbb P}}
\newcommand{\EE}{{\mathbb E}}

\newcommand{\old}[1]{{}}

\title{Hedging our bets:
the expected contribution of species to future phylogenetic diversity}

\author{Mike Steel}
\affil{Biomathematics Research Centre} \affil{University of
Canterbury, Christchurch, New Zealand \\(Corresponding author.
Phone: +64-33667001, Fax: +64-33642587)}
\email{m.steel@math.canterbury.ac.nz}

\author{Aki Mimoto}
\affil{IRMACS}
\affil{Simon Fraser University, Burnaby, Canada}

\and

\author{Arne \O.  Mooers}
\affil{Institute for Advanced Study}
\affil{Berlin, Germany}
\email{amooers@sfu.ca}

\shorttitle{Taxon-specific indices of expected future biodiversity}
\shortauthors{Steel {\em et al.}}

\date{20 April 2007}

\begin{document}

\begin{abstract}
If predictions for species extinctions hold, then the `tree of life'
today may be quite different to that in (say) 100 years. We describe
a technique to quantify how much each species is likely to
contribute to future biodiversity, as measured by its expected
contribution to phylogenetic diversity.  Our approach considers all
possible scenarios for the set of species that will be extant at
some future time, and weights them according to their likelihood
under an independent (but not identical) distribution on species
extinctions. Although the number of extinction scenarios can
typically be very large, we show that there is a simple algorithm
that will quickly compute this index. The method is implemented and
applied to the prosimian primates as a test case, and the associated
species ranking is compared to a related measure (the `Shapley
index'). We describe indices for rooted and unrooted trees, and a
modification that also includes the focal taxon's probability of
extinction, making it directly comparable to some new conservation
metrics.
\end{abstract}

\maketitle

{\em Keywords:}  phylogenetic diversity, extinction, biodiversity
conservation, Shapley index

{\em Short title:}  Taxon-specific indices of expected future
biodiversity
\newpage

\section{Introduction}

Within a given taxonomic group, individual biological species are
generally considered to be of equal or  near-equal biodiversity
value. So, for instance, areas with a greater number of species are
more valuable than those with fewer (Myers et al. 2000). When wild
species are ranked by value, this is usually based on their threat
of extinction (see, e.g. SARA 2002). However, as pointed out by
Cousins (1991), species are discovered and identified because they
are different from other species, which suggests that they may
differ in value. In the context of conservation, Avise (2005) has
highlighted five different currencies for valuing species: rarity,
distribution, ecology, charisma, and phylogeny. Here, we consider
the value of a species based on its position in a phylogeny.  A
phylogeny is the directional, acyclic graph depicting relationships
between leaves (species), which we define formally in the next
section.  A phylogeny generally has a root (which assigns direction)
and edge weights that can represent unique feature diversity (e.g.
as measured by evolutionary time or genetic distance). Species can
be defined by the features they possess, and one measure of their
worth is their expected contribution of unique features. In this
way, we can use a phylogeny to assign a measure of evolutionary
value to a species  based on its expected contribution of unique
features. Because of the  highly imbalanced shape of the Tree of
Life, some species in a phylogeny will have far fewer close
relatives than others in that phylogeny (Mooers and Heard 1997), and
these more distantly-related species will be expected to contribute
more unique features (Faith 1992).

Phylogenetic measures of conservation value have a long pedigree
(see, e.g.  Alschul and Lipman 1990; May 1990) and have begun to be
explored in some detail (Haake et al. 2005; Hartmann and Steel 2007;
Pavoine et al. 2005{\em a}, 2005{\em b}; Redding and Mooers 2006).
So, for example, Pavoine and colleagues presented one new measure of
originality, a set of sampling weights such that the expected
pairwise distance on the tree is maximized. Haake and colleagues
extended the `Shapley value' (Shapley 1953) from co-operative game
theory to the conservation setting to calculate the average distance
of a focal species to all possible subsets of taxa.  For both
measures, more original species are those expected to contribute
more to the resulting sets.  Yet another measure that uniquely
apportions the tree to its tips (Isaac et al. 2007) and which is the
focus of a new international conservation initiative (the EDGE
initiative, Zoological Society of London) scales almost perfectly
with the Shapley value (unpublished results).

One question with these measures concerns the sets that individual
species are asked to complement. For instance, given known
extinction probabilities for species, some future sets are much more
likely than others and so some species will  be more valuable
because their close relatives are less likely to be included in
future sets.  Here we formalize this idea to extend the Shapley
value of a species to include pre-assigned extinction probabilities.
We then compare our measure with the original Shapley value using
the prosimian primates as a test case.

\bigskip

{\bf Definition} Let $\T$ be a rooted or unrooted phylogenetic tree
with leaf set $X$, together with an assignment of positive lengths
to the edges (branches) of $\T$. We let $l(e)$ denote the length of
edge $e$, and let $E(\T)$ denote the set of edges of $\T$. For a
subset $S$ of $X$, let $PD(S)$ denote the {\em phylogenetic
diversity} of $S$ defined as follows. If $\T$ is unrooted then
$PD(S)$ is the sum of the lengths of the edges (branches) of $\T$ in
the minimal subtree that connects $S$. If $\T$ is rooted, then
$PD(S)$ is the sum of the lengths of the edges of $\T$ in the
minimal subtree of $T$ that connects $S$ and the root of the tree.
Figure 1 illustrates these concepts, and includes values at the tips
that we will use in the next section. Note that although the branch lengths in this example
are clock-like, this assumption is not required in any of the results we describe.

\begin{figure}[ht]
\begin{center}
\includegraphics[angle=0,scale=1.5]{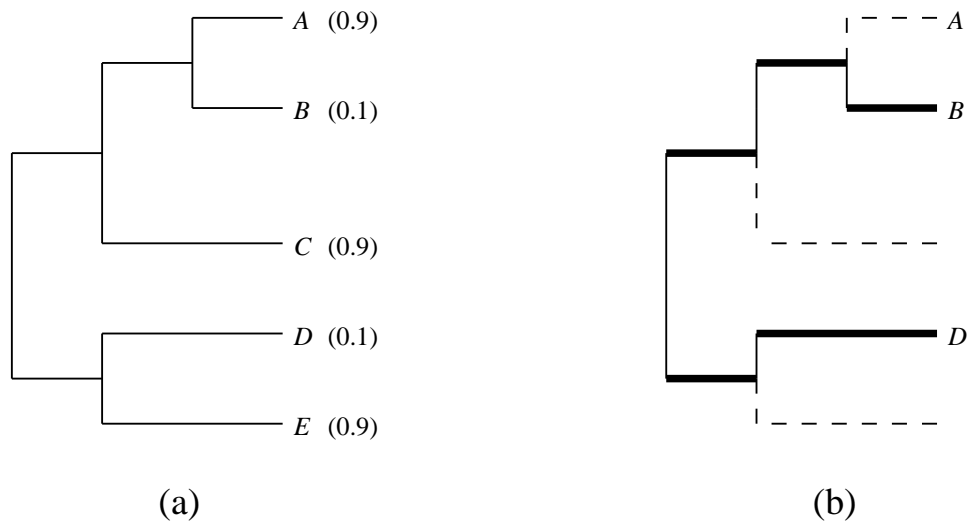}
\caption{(a) A small rooted tree with edge lengths (of $2$ units for
the terminal edges incident with C,D,E, and $1$ unit for the other
five edges). Each tip $j$ has an associated extinction probability
$P(ext)= \epsilon_j$. (b) For a subset $S=\{B,D\}$ of taxa that are
extant at some future time, the phylogenetic diversity score $PD(S)$
is the sum of the lengths of the edges indicated in bold. The dashed
edges lead to extinct taxa.} \label{Fig1}
\end{center}
\end{figure}

\section{The HED index}
  For a  leaf $i \in X$, and a subset $S \subseteq X-\{i\}$ let
$$\Delta_{PD}(S,i): = PD(S \cup \{i\}) - PD(S).$$
The quantity $\Delta_{PD}(S,i)$ measures how much phylogenetic diversity $i$ contributes to the tree that one obtains from $\T$ once species not in $S$ have been pruned out (for example if they go extinct). Alternatively, $\Delta_{PD}(S,i)$ is the marginal increase in phylogenetic diversity of $S$ if $i$ is added.

Now, suppose that each species has an associated extinction
probability $P(ext)$ (which may vary from species to species) ---
for example, this may be the probability that the species is extinct
in (say) 100 years from now (either globally, or in some specified
community).   We will denote this $P(ext)$ value for species $j$ by
$\epsilon_j$.
In this paper we consider the simplest model which
assumes that the extinction of each species in $X$ comprise
independent events.  Given $i \in X$, let $\x_i$ denote the random
subset of species in $X-\{i\}$ which survive (i.e. do not go
extinct).

\noindent By the independence assumption we have:
$$\PP[\x_i = S] = \prod_{j \in S} (1-\epsilon_j) \times \prod_{j \in X-\{i\}-S}
\epsilon_j.$$

\noindent For $i \in X$, let
$\psi_i$ denote the expected value of $\Delta_{PD}(\x_i,i)$.  That is,
\begin{equation}
\label{defeq}
\psi_i = \EE[\Delta_{PD}(\x_i,i)] = \sum_{S \subseteq X-\{i\}} \PP[\x_i=S] \Delta_{PD}(S,i).
\end{equation}
We call $\psi_i$ the {\em heightened evolutionary distinctiveness} of
species $i$, and the function $i \mapsto \psi_i$ the {\em heightened
evolutionary distinctiveness} (HED) index for $\T$. Notice that if all the species in $X-\{i\}$ were guaranteed to survive, then
$\psi_i$ would be just the length of the pendant edge incident with leaf $i$, however random extinctions mean that $\psi_i$ will tend to be
increased (`heightened') over this pendant edge length.

A related but different index, based on the Shapley value in
co-operative game theory, has recently been described by Haake et
al. (2005). This index, denoted here as $\psi^{\rm sh}$ can be
defined (for unrooted trees) as follows: For $i \in X$,
\begin{equation}
\label{defeq2}
\psi^{\rm sh}_i = \frac{1}{|X|}\sum_{S \subseteq X-\{i\}} \binom{|X|-1}{|S|}^{-1} \Delta_{PD}(S,i).
\end{equation}
This index has certain appealing properties. In particular, $\sum_{i
\in X} \psi^{\rm sh}_i = PD(X)$, and there is a simple formula for
quickly computing $\psi^{\rm sh}_i$. The index $\psi^{\rm sh}$ also
has a stochastic interpretation, but this is not based on extinction
or survival of species, rather on the expected contribution to $PD$
of each species under all possible orderings of the total set of
species (for details see Haake et al. 2005). The index $\psi^{\rm
sh}$ allocates existing $PD$ `fairly' amongst the species, whereas
$\psi$ quantifies the expected  contribution of each species to
future $PD$.

\section{Computing the HED index}
Computing the HED index directly via (\ref{defeq}) could be
problematic as it requires summation over all the subsets of
$X-\{i\}$ and this grows exponentially with $|X|$.  However we now
show that the index can be readily and quickly computed, both for
rooted and unrooted trees.  This polynomial-time algorithm for
computing $\psi$ thus  complements (but is quite different to) the
polynomial-time algorithm described by Haake et al. (2005) for
computing $\psi^{\rm sh}$.

\subsection{Rooted trees}
For a rooted phylogenetic $X$--tree $\T$ let $C(e)$ denote the set
of species in $X$ that are descended from $e$ (i.e. the clade that
results from deleting $e$ from $\T$). For $i \in X$, let $e_1, e_2,
..., e_k$ ($k = k(i) \geq 1$) denote the edges (branches) on  the
path from $i$ to the root of $\T$, listed in the order they are
visited by that path. Recall that $l(e)$ denotes the length of edge
$e$. The proof of the following theorem is given in the Appendix.

\begin{theorem}
\label{th1}
$$\psi_i = \sum_{r =1}^k l(e_r)\prod_{j \in C(e_r)-\{i\}}\epsilon_j$$
\end{theorem}
Note that in this (and the next) theorem we adopt the convention $\prod_{j \in \emptyset} \epsilon_j =1$, which is relevant for the first term ($r=1$) in the sum as $C(e_r)-\{i\}$ is empty. Thus the first term in the summation expression for $\psi_i$ given by Theorem~\ref{th1} is simply $l(e_1)$, the length of the pendant edge of $\T$ incident with species $i$.

\subsection{Example}
We can apply the HED index to the members of the rooted tree
depicted in Fig. 1.  For example, to compute $\psi_A$ by using
Theorem~\ref{th1} we have $\psi_A = 1+ 1\cdot\epsilon_B + 1\cdot
\epsilon_B\epsilon_C = 1.19$. By inspection, we can see that the most valuable species will be $D$, since it shares an
edge with only one other species above the root, and that this species ($E$) has a high
$P(ext)$. At the other extreme, $A$ shares its path to the root with
two other species, and one of them ($C$) has a low $P(ext)$. It
should therefore receive a low HED value.  The computed values are
$\psi_D=2.9$, $\psi_B=2.71$, $\psi_E=2.1$, $\psi_C=2.09$, and
$\psi_A=1.19$. Using the Shapley index (Haake et al. 2005), $D$ and
$E$ are ranked first (with value = 2.63), followed by $C$ (2.33) and
then $A$ and $B$ (1.75).  Pavoine's  QE metric (Pavoine et al. 2005)
returns the same ranking as does the Shapley. A portal for computing
HED is available at http://www.disconti.nu/-phylo/emd.dpf

\subsection{Unrooted trees, and properties of the index}
We now provide a similar formula for efficiently computing the HED index for unrooted trees.
Given a leaf $i$ of $\T$ and an edge $e$ of $\T$, $e$ induces a split of $X$ into two disjoint subsets,  and one of these subsets, which we
denote as $C_i(e)$, contains $i$. The proof of the following theorem is given in the Appendix.

\begin{theorem}
\label{th2}
$$\psi_i = \sum_{e \in E(\T)} l(e) \cdot (\prod_{j \in C_i(e)-\{i\}} \epsilon_j ) \cdot (1-\prod_{j \in X- C_i(e)}\epsilon_j)$$
\end{theorem}

Notice that the rooted HED index is just a special case of the unrooted HED index (indeed Theorem~\ref{th1} can be deduced from Theorem~\ref{th2}). To see this, given a rooted tree $\T$ attach a new leaf $\rho$ to the root via a new edge to obtain an unrooted tree, and assign the new edge length $0$. Let $\epsilon_{\rho}=0$.  Then it is easily seen that the HED index for $\T$ is just the HED index for
the derived unrooted tree.

Using Theorem~\ref{th1} it can be shown that if $\T$ is a rooted phylogenetic  tree
then the condition:
\begin{equation}
\sum_{i \in X} \psi_i = PD(X),
\label{pareto}
\end{equation}
holds for all selections of positive branch lengths and $\epsilon$'s
if and only if $\T$ is a `star tree' (that is, every leaf is
adjacent to the root). Moreover Theorem~\ref{th2} shows that there
is no unrooted phylogenetic tree $\T$ for which (\ref{pareto}) holds
for all positive branch lengths and $\epsilon$ values (of course
(\ref{pareto}) may hold on phylogenetic trees -- either rooted or
unrooted -- if the branch lengths and $\epsilon$ values take certain
values). This contrasts with  the index $\psi^{\rm sh}$ which
satisfies $\sum_{i \in X} \psi_i^{\rm sh} = PD(X)$ on all unrooted
phylogenetic trees and choices of branch lengths, a property that is
referred to as the Pareto efficiency axiom by Haake et al. (2005).
In the setting of this paper we should not be surprised that
(\ref{pareto}) holds for $\psi$ only in very special cases since we
are not trying to divide out existing $PD$ amongst present taxa (one
motivation behind $\psi^{\rm sh}$) but rather quantify the expected
contribution each species makes to future $PD$.

\section{Application}

We compared the HED index with the Shapley (Haake et al. 2005)
values for the Prosimians (Mammalia: Primata), a group of
approximately 50 species with a broad range of extinction
probabilities.  This group includes the Aye-Aye, the lemurs, the
lorises and galagos.  We made use of a recent dated Supertree of the
order Primates (Vos and Mooers 2004; Vos 2006), see Fig. 2, and Red
List risk designations from the IUCN (www.iucnredlist.org, accessed
February 2006). Following Isaac et al. (2007) and Redding and Mooers
(2006), we first converted the five categories of risk (CR, EN, VU,
NT, and LC) to probabilities of extinction. Under the IUCN criteria,
the species in the VU category are given a $P(ext)$=0.1 over the
next 100 years. We gave the lowest and highest threat categories
very conservative probabilities of extinction over the next 100
years of 0.001 and 0.9 respectively, leaving $P(ext)=0.5$ for  EN,
and $P(ext)=0.01$ for NT: this scale is very similar to that
calculated from real population viability analyses for birds
(Redding and Mooers 2006). We are primarily interested in how the
ranking of species changes using different approaches.

\begin{figure}
\begin{center}
\includegraphics[angle=0,scale=.70]{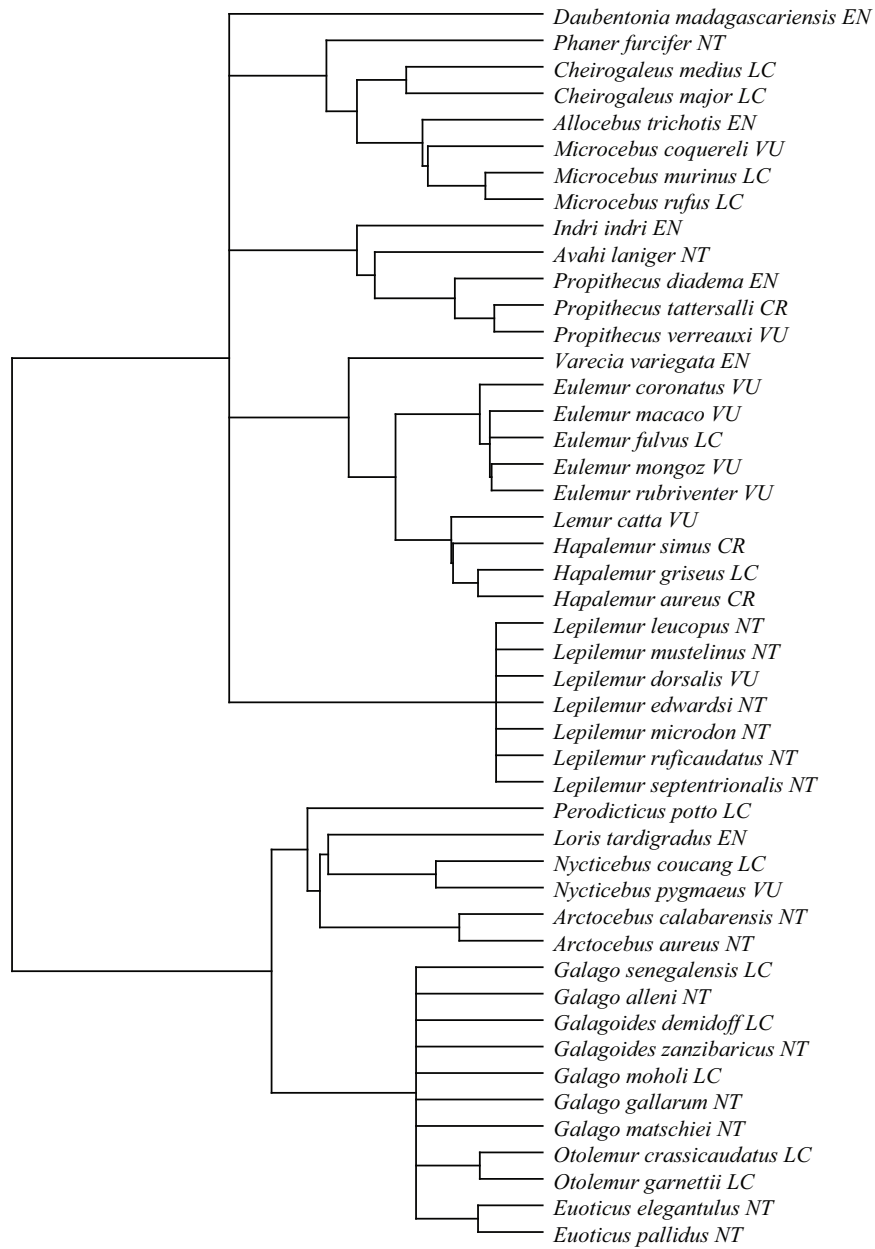}
\caption{Prosimian species tree and associated IUCN threat categories. $CR$: critically endangered, $P(ext)=0.9$; $EN$: endangered, $P(ext)=0.5$; $VU$: vulnerable, $P(ext)=0.1$; $NT$: near threatened, $P(ext)=0.01$; $LC$: least concern, $P(ext)=0.001$.  Edge lengths are on an arbitrary scale that represents time since divergence.}
\label{Fig2}
\end{center}
\end{figure}

The bivariate correlation between the metrics is high (0.94).  Both measures chose the Aye-Aye  ($Daubentonia$ $madagascarensis$) as the most important species, followed by $Perodicticus$ $potto$.   Interestingly, the three most highly ranked species under current conservation policy (the critically endangered lemurs $Propithecus$ $tattersalli$, $Hapalemur$ $simus$, $H.$ $aureus$) are nested well up in the tree (Figure 2) such that none of them were chosen in the top ten for either SV or HED. If we compare the rest of the rankings for these two metrics, the
largest single difference is for  the two $Arctocebus$ species:  they rank twelfth under
SV (being relatively isolated on the tree), but only twenty-sixth under HED:  because
neither is severely threatened,  the chances are good that their common path will persist.

Both measures are very heavily influenced by the pendant edge ($PE$) length of the focal species (with correlations $PE$ vs. SV=0.94, and $PE$ vs. HED=0.98).  $PE_{i}$  is always part of the marginal increase to $S$, while interior nodes are most likely represented with high probability, especially for larger and more balanced trees.  $PE$ is, however, a poor predictor of HED for  $Propithecus$, for  $H.$ $griseus$, and for $Nycticebus$ $couang$ (Figure 2).   The first two groups contain the three most endangered species, increasing the value of close relatives.  $Nycticebus$ is an isolated genus, and $N. couang$'s sister species is listed as vulnerable ( $P(ext)=0.1$).  Likewise for $Propithecus verreauxi$ - although it has close relatives and so a short PE,
these relatives are at high risk of extinction, which increases its value; this is what
we saw with species $D$ in figure 1.

\subsection{Incorporating the focal taxon's extinction risk (HEDGE scores)}

The effect of close relatives' risk status on one's own value is
precisely the strength of the HED approach.  However, the fact that the
extinction risk of other species affects a focal species, but its own
risk does not is somewhat counter-intuitive. We address this by showing how it is possible to write the HED index as the sum of two terms each of which takes into account the extinction risk of the focal species.
To describe this further, let $I$ be the random variable which takes the value $i$ if the focal species $i$ survives (at the future time under consideration) and which
otherwise takes the value of the emptyset (i.e. $\emptyset$) if $i$ goes extinct.

\noindent Let $$\psi_i' = \EE[PD(\x_i \cup \{i\}) - PD(\x_i \cup I)],$$
where, as before $\x_i$ is the random subset of species in $X-\{i\}$ that survive.
In words, $\psi_i'$ is  the increase in the expected PD score if we condition on the event that species $i$
survives.

\noindent Similarly, let $$\psi_i'' =  \EE[PD(\x_i \cup I) - PD(\x_i)].$$
In words, $\psi_i''$ is  the decrease in the expected PD score if we condition on the event that species $i$
becomes extinct. The following result describes how to compute these two indices easily from the HED index, and verifies that they add together to give the HED index (its proof is given in the Appendix).
\begin{theorem}
\label{th3}
\mbox{}
\begin{itemize}
\item[{\rm (i)}]
$\psi_i' = \epsilon_i\cdot\psi_i$,
\item[{\rm (ii)}]
$\psi_i'' = (1-\epsilon_i)\cdot\psi_i$,
\item[{\rm (iii)}]
$\psi_i' + \psi_i'' = \psi_i$.
\end{itemize}
\end{theorem}

The approach of assigning a value to a species which is a function
of its phylogenetic distinctiveness and its extinction probability
has been referred to as `expected loss' by Redding and Mooers (2006)
and, more evocatively, an `EDGE' score (Evolutionarily Distinct and
Globally Endangered) by Isaac et al. (2007).

In the same spirit we will call $\psi'$ and $\psi''$ (which extend
our HED index $\psi$) HEDGE ({\em heightened evolutionary
distinctiveness and globally endangered}) scores. The HEDGE score
$\psi_i'$ is more relevant when evaluating actions that might save
species, whereas the HEDGE score $\psi_i''$ is appropriate when
evaluating actions that might cause the extinction of species (such
as building a dam).

One potential advantage of HED and HEDGE over previous scores is their flexibility in
designing conservation scenarios. So for
instance, we can choose IUCN-ranked species for which conservation
is cheap and/or already partially successful, set their $P(ext)$  to
0, and see how rankings of other species change. Alternatively, we
might want to increase the $P(ext)$ to 1.0 for certain species to
see how others are affected.

Most generally, HED and HEDGE could be incorporated in an assessment
of species value that included many factors besides risk and future
contribution, e.g. the ecological, distributional and aesthetic
values enumerated by Avise (2005), and the costs of recovery and
probability of its success.

\newpage
\section{Appendix: Proofs of theorems.}

\noindent{\em Proof of Theorem~\ref{th1}}

\noindent First observe that the only edge lengths that contribute to
$\Delta_{PD}(S,i)$ are those from the set $\{l(e_1), l(e_2), \ldots, l(e_k)\}$.

\noindent Consequently, for the random set $\x_i$ of surviving species of $X-\{i\}$ we have
 $$\Delta_{PD}(\x_i,i) = \sum_{r\geq 1} l(e_r) \cdot I_r(\x_i)$$
where
 $I_r(\x_i)$ is
the $0,1$ indicator random variable that takes the value $1$ precisely
if $e_r$ is not an edge of the subtree of $\T$ connecting the taxa in
$\x_i$ and the root of $\T$; since this is the only situation for which
$e_r$ lies in the subtree of $\T$ connecting $\x_i \cup \{i\}$ but not in
the subtree of $\T$ connecting $\x_i$.

\noindent Thus, by linearity of expectation,
$$\psi_i = \EE[\Delta_{PD}(\x_i,i)] = \EE[\sum_{r\geq 1} l(e_r) \cdot I_r(\x_i)] = \sum_{r\geq 1} l(e_r) \cdot \EE[I_r(\x_i)],$$
and since $I_r(\x_i)$ takes the values $0$ and $1$, $\EE[I_r(\x_i)]
= \PP[I_r(\x_i)=1]$. Thus,
$$\psi_i = \sum_{r \geq 1}l(e_r) \cdot \PP[I_r(\x_i)=1].$$
Now, the event `$I_r(\x_i)=1$' occurs precisely if none of the
elements in $C(e_r)-\{i\}$ survive, and this latter event has
probability $\prod_{j \in C(e_r)-\{i\}}\epsilon_j$.  Substituting
this into the previous equation establishes the theorem.
\hfill$\Box$

\noindent{\em Proof of Theorem~\ref{th2}}

\noindent For $i \in X$ and the random subset $\x_i \subseteq X-\{i\}$, we have
$$\Delta_{PD}(\x_i,i) = \sum_{e \in E(\T)} l(e) \cdot I_e(\x_i)$$
where $I_e(\x_i)$ is the $0,1$ indicator random variable taking the value $1$ precisely if $\x_i$ consists of no elements of $
C_i(e)-\{i\}$ and at least one element of $X-C_i(e)$.

\noindent Thus,
$$\psi_i = \sum_{e \in E(\T)} l(e) \cdot \PP[I_e(\x_i) =1]$$
and by the independence assumption $$\PP[I_e(\x_i) =1] = \PP[\x_i \cap (C_i(e)-\{i\}) = \emptyset]\cdot  \PP[\x_i \cap (X- C_i(e)) \neq  \emptyset]$$ and so
$$\PP[I_e(\x_i) =1] =
(\prod_{j \in C_i(e)-\{i\}}\epsilon_j ) \cdot (1-\prod_{j \in X- C_i(e)}\epsilon_j),$$
as claimed.
\hfill$\Box$

\noindent{\em Proof of Theorem~\ref{th3}}

\noindent By definition $\psi_i = \EE[PD(\x_i \cup \{i\}) - PD(\x_i)]$, and so $\EE[PD(\x_i \cup \{i\})] = \psi_i +\EE[PD(\x_i)]$.

\noindent Now we can write the unconditional expectation  $\EE[PD(\x_i \cup I)]$ as the weighted sum of conditional expectations
$\EE[PD(\x_i \cup I)|I =\{i\}] \PP(I = \{i\}) + \EE[PD(\x_i \cup I)|I =\emptyset] \PP(I = \emptyset)$ and so
$$\EE[PD(\x_i \cup I)] =(1-\epsilon_i)\EE[PD(\x_i\cup \{i\})] + \epsilon_i \EE[PD(\x_i)].$$

\noindent Parts (i) and (ii) now follow by applying these equations (and the linearity of expectation) to the definitions of $\psi_i'$ and $\psi_i''$.
Part (iii) follows directly from parts (i) and (ii).
\hfill$\Box$

\section{Acknowledgments}
This research was supported by a grant from the Marsden Fund, New Zealand to MS and AOM.
AOM received additional support from NSERC Canada and the Institute for Advanced Study,
Berlin. We thank Dave Redding, Klaas Hartmann, and Andy Purvis for discussion.

\end{document}